\documentclass[12pt,a4paper]{article}
\usepackage{amsfonts}
\usepackage{color}
\usepackage{amssymb}
\usepackage{amsmath}
\usepackage{graphicx,color}
\usepackage{amsthm}
\usepackage{enumerate}
\usepackage{bm}
\usepackage{fancyhdr}
\usepackage{pdfsync}

\newtheorem*{obs}{Observation}

\newtheorem{lemma}{Lemma}
\newtheorem*{lemma*}{Lema}

\numberwithin{equation}{section}

\newcommand{\comp}{\mathbb{C}}

\newcommand{\mpro}{\mathcal{M}}

\newcommand{\upro}{\mathcal{U}}

\newcommand{\I}{\mathbb{I}}

\newcommand{\R}{\mathbb{R}}
\newcommand{\N}{\mathbb{N}}

\newcommand{\ldos}[2]{{L}^2(#1 , #2)}

\newcommand{\rdos}[2]{{H}^{-\frac{1}{2}}(#1 , #2)}

\newcommand{\tr}{{\rm tr}}
\newcommand{\e}{{\rm e}}

\newcommand{\ol}{{\rm {\bf L}}}
\newcommand{\lap}{{\rm {\bf \Delta}}}
\newcommand{\oh}{{\rm {\bf H}}}

\date{\small \today}
\begin{document}

\title{\bf Quantum Fields bounded by one dimensional crystal plates}

\author{
J.M. Mu$\tilde{\rm n}$oz-Casta$\tilde{\rm n}$eda\footnote{jose.munoz-castaneda@uni-leipzig.de}
$~${\small and}
M. Bordag\footnote{bordag@itp.uni-leipzig.de}\\
\footnotesize{\sl Institut f\"ur Theoretische Physik, Universit\"at Leipzig, Germany}\hfill}
\maketitle

\begin{abstract}
  We get deeper understanding of the role played by boundary conditions in quantum field theory, by studying the structure of a scalar massless quantum field theory bounded by two one dimensional planar crystal plates. The system can also be understood as a massless scalar confined to propagate in the surface of a finite cylinder. We classify the most general type of regular behaved boundary conditions that the quantum field can satisfy, in accordance with the unitarity principle of quantum field theory. Also, we characterize the frequency spectrum for each quantum field theory, by computing the holomorphic spectral function. The spectral function is the starting point to compute the Casimir energy as a global function over the space of allowed boundary conditions for the quantum field theory.
\end{abstract}

\section{Introduction}
Boundary conditions play an important role in quantum field theory. On the one hand side they are generalizations of interactions which are concentrated on a region of co-dimension 1 (or higher). Typical examples are conductor boundary conditions which are a good idealization if the skin depth drops below the wavelength of the relevant electromagnetic field modes \cite{Jackson75}. On the other hand, periodic boundary conditions appear naturally when considering topological objects, for example a quantum field on a torus, or quantum field theory at finite temperature. Other examples for the use of boundary conditions can be found   in the quantum Hall effect \cite{halperin82,stonemathew}, in the physics of graphene \cite{Berry:1987qi,Peres:2006zz}, in planar plasma systems (see references \cite{fett73-81-367,BORDAG2007B}) etc. Boundary phenomena are rather relevant for the structure of the vacuum and the low energy spectrum than for high energy phenomena. For field theories without mass scale, i. e. conformal theories, these effects are amplified because the long distance correlations allow these boundary effects to be observed over all the interior region. Precisely, the conformal field theories related to statistical models with second order phase transitions and string theories, are the ones where the boundary effects have more remarkable applications. Most of the physical systems appearing in nature or implemented on a laboratory, have very large but finite physical dimensions. This makes of great interest the study of quantum fields confined in bounded domains.
\par
A more mathematical question is that about the most general boundary conditions compatible with the general principles of self-adjointness of the Hamiltonian and unitarity. These questions are discussed, for example, in \cite{jmmcphd,Asorey:2004kk} in detail. It is worth mentioning that these boundary conditions are in general non-local and can encode essential physical information to a large extend. In other words, many physical problems can be formulated in terms of boundary conditions. From a more mathematical point of view the boundary conditions are used to obtain ensure discrete spectrum of linear operators in Hilbert spaces \cite{dunford2}.

This paper is focused in the applications, of a systematic formalism for boundary conditions in quantum
field theory \cite{jmmcphd}, to the Casimir effect. Since the seminal paper of H. B. G. Casimir \cite{Casimir:1948dh}, where the effect was theoretically proposed, and the first experimental measurements in 1958 by Sparnaay \cite{SPARNAAY1958}, many theoretical and experimental work have been done (see references \cite{bordagbook,klim09-81-1827,miltonbook}). Since the theoretical and subsequent experimental discovery of the Casimir effect, theoretical research on Casimir effect has been focused on two separated directions. On one hand, the study of the dependence of the Casimir energy on the geometry of the objects subjected to the vacuum force. In this area are specially remarkable the references \cite{Emig:2007cf,WIRZBA2006} giving for the first time a general method that allows to calculate (numerically or analytically) the vacuum energy between compact objects of arbitrary shape like spheres and cylinders which is free of UV divergences. On the other hand, instead of studying the dependence of the vacuum energy on the geometry one may study the dependence on the boundary
conditions that the quantum fields satisfy at the surfaces of interacting objets. Actually, related physical interest is the influence of corrugations on the Casimir force (see, for instance, in the review \cite{klim09-81-1827}). Especially periodic corrugations can be mapped onto non-trivial boundary conditions keeping periodicity \cite{Golestanian:1998bx}.
Much work was done on periodic gratings, \cite{lamb08-101-160403,chiu10-81-115417}
More general methods developed in reference \cite{Emig:2007cf} allow to implement some families of boundary conditions in the path integral approach. Also reference \cite{bordagbook} shows how to introduce local boundary conditions on the quantum fields in the path integral approach. But the path integral approach does not allow to implement the most general type of boundary conditions that the quantum fields can satisfy (see references \cite{Asorey:2007zza,Asorey:2006ij}), and especially problems when implementing Neumann boundary conditions are discussed in \cite{grosche95-jpa}).
Similar problems appear in the approach \cite{gies06-74-045002} using world-line methods.

In the last years, the advances in experimental nano-scale physics allowed to implement a huge family of exotic
boundary behavior, producing results of great importance in the Casimir effect. M. Asorey, J. M. Munoz-Castaneda {\it et al} developed in the last years a theoretical framework that classifies the boundary conditions that quantum fields can satisfy, as well as allows to study the dependence of the Casimir energy on the most general possible type of allowed boundary conditions (see references \cite{jmmcphd,Asorey:2006pr,Asorey:2007rt,Asorey:2007kw,Asorey:2007zza,Asorey:2008xt}). The geometry of the
bounded system in this case is a simple geometry, equivalent to the original geometry studied by Casimir
(parallel infinite plates), but the boundary conditions are most general ones. In the case of infinite
parallel homogeneous isotropic plates Asorey, Munoz-Castaneda {\it et al} classified all
the allowed boundary conditions for the quantum fields attending to the type of Casimir force they give
rise to: attractive, repulsive or null. Focused on the Casimir energy calculation, the central object in
Asorey-Munoz-Castaneda (AMC) formalism is the spectral function. The spectral function is defined as the
function whose zeros are the frequencies of the quantum field theoretical hamiltonian of the system. Calculation
of this function allowed them to compute the value of the Casimir energy for the most general type of boundary
condition, and hence to study the Casimir energy as a function over the space of allowed boundary conditions.
\par
The purpose of this paper, is to use AMC formalism to calculate the space of boundary conditions in a more
complicated case, and then give an explicit expression for the corresponding spectral function, in terms
of the boundary condition. This calculation will provide the machinery necessary to compute the corresponding
Casimir energy as a function over the space of boundary conditions that the quantum fields can satisfy.
The case to be studied is the case of two parallel non homogenous plates (1 dimensional crystal plates),
isomorphic to a cylindrical geometry. Via this calculation we try to get a deeper understanding of the behavior of the vacuum interaction in crystallographic systems taking into account the microscopic lattice structure. In order to take into account the lattice structure we will impose a periodical behavior over the scalar field that gives rise to the vacuum interaction. This interaction can be also understood as a boundary effect due to the finiteness of a 2 dimensional crystals in one of its directions.
\\Throughout this paper we put $\hbar=c=1$.

\section{The Quantum scalar field in bounded domains}
Let $\R\times M\subset\R^{D+1}$ be a space-time, where the spatial surface $M\in\R^D$ is a
bounded domain with regular boundary $\partial M$. The dynamics of a free complex scalar
field propagating in $\R\times M$ is given by the action
\begin{equation}
  S(\phi)={1\over 2}\int_{_{\R\times M}}d^{^{D+1}}x\left( \partial^\mu\phi^*\partial_\mu\phi+m^2|\phi|^2\right)
  -{1\over 2}\int_{_{\R\times\partial M}}d^{^{D}}x\phi^*\partial_n\phi,
\end{equation}
where $\partial_n$ denotes the outgoing normal derivative. The boundary term is added in
order to obtain the usual free field equations of motion in the bulk without any extra boundary term potential
\begin{equation}
  -\partial_t^2\phi=\ol\phi,\quad\ol=-\lap+m^2,
\end{equation}
where $\lap$ is the Laplace operator induced in the bounded domain $M$ by the Euclidean structure of $\R^D$.
\par
The usual canonical quantization procedure leads to the quantum Hamiltonian
\begin{equation}
  \oh={1\over2}\int_{_{M}}d^{^{D}}x\left(\widehat{\pi}(x)^*\widehat{\pi}(x)+\widehat{\phi}(x)^*\ol
  \widehat{\phi}(x)\right)
\end{equation}
where the canonical momentum density operator is given by $\widehat\pi(x)={1\over 2}\partial_t\phi^*(x)$, and satisfies the canonical commutation rules $\left[\widehat{\pi}(x),\,\widehat{\phi}(x')\right] =-i\delta(x-x')$
\par
Decomposing the fields in eigen-modes  of $\ol$,
\begin{equation}
  \ol\phi_k(x)=\lambda_k\phi_k(x).
\end{equation}
allows to write the system as an infinite collection of non-coupled harmonic oscillators. In order to have a unitary quantum field theory we should require the eigen-values $\{\lambda_k\}$ to be real (selfadjointness of $\ol$), and non negative (non-negativity of operator $\ol$). When $M=\R^D$, then automatially $\ol$ is self-adjoint and non-negative because $\lap$ is self-adjoint and positive definite in $\R^D$. But when $M\subset\R^D$ is a bounded domain with regular boundary, $\lap$ may not be essentially self-adjoint\footnote{A symmetric operator ${\cal O}$ defined on a dense domain of a Hilbert space ${\cal H}$ is said to be essentially self-adjoint if it admits an unique self-adjoint extension (for more details on the theory of self-adjoint operators and self-adjoint extensions see reference \cite{dunford2}).} and in that case one needs to consider an infinite set of self-adjoint extensions (\cite{Asorey:2004kk,jmmcphd}), which are in general not definite positive (\cite{Asorey:2004kk,jmmcphd}).

\subsection{AIM formalism for self-adjoint extensions.}
To ensure self-adjointness of operator $\ol$ we might use the Asorey-Ibort-Marmo (AIM) formalism for self
adjoint extensions\footnote{The AIM formalism for self-adjoint extensions is equivalent to the classical formalism for self-adjoint extensions of symmetric operators with dense domains in Hilbert spaces developed by Von Neumann (see ref. \cite{dunford2} for modern description) in terms of defect spaces. The advantage of AIM formalism in physics is that the theory is formulated just in terms of boundary values (and normal derivatives) of the functions that belong to the domain of the self-adjoint extension. The boundary data have a clear physical interpretation, as the boundary data of wave functions (see \cite{Asorey:2004kk}).} in quantum mechanics \cite{Asorey:2004kk}. We will study quantum mechanics in the domain $M$:
\begin{itemize}
  \item Quantum physical states: $\ldos{M}{\comp^N}$.
  \item Quantum Hamiltonian: $\ol=-\lap+m^2\,\,\Rightarrow$ symmetric operator in $C_0^\infty\left(M,\comp^N\right)$
  (smooth functions with compact support in $\stackrel{\circ}M$), but not essentially self-adjoint
  in $\ldos{M}{\comp^N}$.
\end{itemize}
The obstruction for $\ol$ to be self-adjoint is given by the boundary term obtained integrating by parts:
\begin{equation}
  \langle\psi_1,\,\ol\psi_2\rangle=\int_{_{M}}d^{^{D}}x\psi_1^\dagger(\ol\psi_2)=
  \int_{_{M}}d^{^{D}}x(\ol\psi_1)^\dagger\psi_2+i\Sigma(\psi_1,\,\psi_2).
\end{equation}
\begin{equation}
  \Sigma(\psi_1,\,\psi_2)\equiv i\int_{_{\partial M}}\left[(\dot\varphi_1,\,\varphi_2)-
  (\varphi_1,\,\dot\varphi_2)\right]d\mu_{_{\partial M}}
\end{equation}
being $\varphi\equiv\left.\psi\right|_{_{\partial M}}$ and $\dot\varphi\equiv\left.
\partial_n\psi\right|_{_{\partial M}}$. The boundary term $\Sigma(\psi_1,\,\psi_2)$ can be used to characterize the domain where $\ol$ is self-adjoint \cite{Asorey:2004kk}
\begin{center}
  $\bullet$ $\ol$ will be self-adjoint $\Leftrightarrow$ $\Sigma(\psi_1,\,\psi_2)=0$.
\end{center}
For simplicity reasons in this first approach, we will assume that the boundary data lie in
$\ldos{\partial M}{\comp}$. This simplification concerns also to the physical behavior of the boundary,
so it must be carefully explained. In the most general approach $\varphi\in\rdos{\partial M}{\comp}$ and
$\dot\varphi\in\rdos{\partial M}{\comp}\oplus\left.\ker (\lap^\dagger)\right|_{\partial M}$, to encode all
the self-adjoint extensions of $\lap$ (for detailed technical description see chapter 1 in \cite{jmmcphd}).
This restriction must be taken into account in order to allow non-regular behavior of the boundary data
originated by physical situations that might include singular charge distributions on the boundary, or
some kind of singular impurities distributions over the crystal plates. In order to simplify a first
approach to the study of the vacuum interaction between one-dimensional crystal plates, we will avoid
such kind of behaviors, and assume for now on that all the boundary data lie in $\ldos{\partial M}{\comp^N}$,
as was originally proposed in reference \cite{Asorey:2004kk}\footnote{To include non-regular behavior of the boundary
data in the formalism of quantum fields over bounded domains with regular boundary, and the re-formulation
of the two AIM theorems see reference \cite{jmmcphd}.}. Hence in this first approach we will only deal with
those boundary conditions that do not encode physically singular behaviors of the boundaries. We will call
this set of boundary conditions, the set of regular self-adjoint extensions (${\cal M}_{r}$).
\par
The {\bf first AIM theorem} ensures that:
\begin{itemize}
  \item Self-adjoint extensions of $\ol$ are in one-to-one correspondence with the
  unitary group $\upro\left(\ldos{\partial M}{\comp^N}\right)=\mpro_{r}$ of unitary operators
  over the square integrable functions over the boundary of the space.
  \item Given a unitary operator $U\in\mpro_r$ the associated self adjoint extension $\ol_U$
  is characterized by the domain of functions that satisfy the boundary condition
          \begin{equation}
            \varphi-i\dot\varphi=U\left(\varphi+i\dot\varphi\right)\label{gbc}
          \end{equation}
  where now, $\varphi,\,\dot\varphi\in\ldos{M}{\comp^N}$, and $U\in\mathcal{U}\left(\ldos{M}{\comp^N}\right)$
\end{itemize}
Some examples of well known boundary conditions written in this formalism:
\begin{itemize}
  \item Dirichlet boundary condition: $U_d=-\I$.
  \item Neumann boundary condition: $U_n=\I$.
\end{itemize}
When $\pm 1\not\in\sigma(U)$ unitary operator $U$ admits non-singular Cayley transform, and boundary
condition in this case can be rewritten in a more simpler way
\begin{equation}
  \dot\varphi=A\varphi\quad;\quad\varphi=A\dot\varphi
\end{equation}
\begin{equation}
  A=-i\,\frac{\I-U}{\I+U}\quad;\quad A^{-1}=i\,\frac{\I+U}{\I-U}
\end{equation}
where $A$ is a self-adjoint operator in the domain of the corresponding self-adjoint extension. Cayley transformation takes unitary operators ($\pm 1\not\in\sigma(U)$) into self-adjoint operators.
\par
Now we should ensure the positivity of the self-adjoint extension $\ol_U$. This condition is described
by the {\bf Second AIM theorem}, that ensures the existence of unitary operators $U\in\mpro$ for which
$\ol_U$ has negative energy eigen-states (also the theorem characterizes these operators):
\begin{itemize}
  \item Given $U\in\mpro_r$ with $-1\in\sigma(U)$, there are near $U$ uni-parametric families of
  unitary operators which give rise to self-adjoint extensions of $\ol$ with arbitrary negative
  energy states.
\end{itemize}
Now with these two results we should go back to the QFT and compute which of those $U\in\mpro_r$ give
rise to self-adjoint extensions $\ol_U$ that are consistent with the principles of quantum field theory.

\section{Consistency conditions in QFT.}
As we saw in first two sections, a massless scalar QFT over the bounded domain $M$ will be consistent if $\ol_U$
is a non negative self-adjoint extension of $\ol$. By {\bf first AIM theorem} it is ensured that $\ol_U$
is self-adjoint for any $U\in\mpro_r$. It is easy to check from the expression for the boundary condition that:
\begin{equation}
  \langle\phi,\ol_U\phi\rangle=\|\overrightarrow{\nabla}\phi\|^2+m^2\|\phi\|^2+i\left\langle\varphi,\, \frac{\I-U}{\I+U}\varphi\right\rangle\label{vesp}
\end{equation}
By {\bf second AIM theorem} there exist some unitary operators $U\in\mpro$ for which (\ref{vesp}) is negative,
so not any $U\in\mpro_r$ is consistent with the principles of quantum field theory, even though when $\ol_U$
defines a consistent quantum mechanical system. Now observe that in right hand of equation (\ref{vesp}):
\begin{itemize}
  \item $\|\overrightarrow{\nabla}\phi\|^2\geq0$
  \item $m^2\|\phi\|^2\geq0$
  \item $\Rightarrow\quad\|\overrightarrow{\nabla}\phi\|^2+m^2\|\phi\|^2\geq0$
\end{itemize}
{\bf Strong Consistency Condition:} we will call strongly consistent boundary conditions or
consistent boundary conditions, those boundary conditions which give rise to a consistent QFT for any
size $\delta$ of the edge, i. e. those given by unitary operators $U\in\mpro_r$ satisfying
\begin{equation}
  i\left\langle\varphi,\,\frac{\I-U}{\I+U}\,\varphi\right\rangle\geq0,\quad\forall\quad
  \varphi\in\ldos{\partial M}{\comp^N}
\end{equation}
That is to say that strongly consistent boundary conditions are only those characterized
by unitary operators $U\in\mpro_r$ that satisfy
\begin{equation}
  i\,\frac{\I-U}{\I+U}\geq0\label{opcond}
\end{equation}
The consistency condition can be written more explicitly in terms of the eigenvalues of the
operator $U$. Since $U$ is unitary its spectrum is a set of uni-modular complex numbers $\sigma(U)=\{\e^{i\theta}\}$, and taking the diagonal terms in the operator inequality given by (\ref{opcond}), we get a condition over the eigenvalues:
\begin{equation}
  i\,\frac{1-\e^{i\theta}}{1+\e^{i\theta}}\geq0\quad\Longleftrightarrow
  \quad\tan\left({\theta\over 2}\right)\geq0\label{ccond}
\end{equation}
Finally, with the condition (\ref{ccond}) we can specify the space $\mpro_{rF}$,
\begin{equation}
  \mpro_{rF}=\left\{U\in\mpro\,|\,\forall\,\lambda=\e^{i\theta}\in\sigma(U),\,
  \tan\left({\theta\over 2}\right)\geq0\right\},\label{mf}
\end{equation}
which is the space  of consistent boundary conditions for the QFT.

\section{2+1 QFT: 1-dimensional crystal plates.}
\begin{figure}[htbp]
  \centerline{\includegraphics[height=6.5cm]{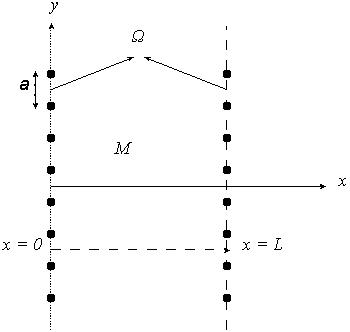}}
  \caption{\footnotesize{The physical space limited by two one dimensional lattices.}} \label{1dwire}
\end{figure}

The analogue of the 1+1 dimensional example explained above is the case of a 2+1 dimensional quantum
field theory, in a bounded space limited by two infinite homogeneous isotropic parallel wires. This case
and its generalizations to general D+1 scalar quantum field theories has been studied in detail in
\cite{jmmcphd}. The simplest extension that can be done for all these plate geometries is to broke the
isotropy of the plates, by introducing a D-1 dimensional lattice over each plate. To learn how the boundary
conditions behave in this case the simplest case we can study is that one given by two parallel and identical
crystal wires, as is shown in figure \ref{1dwire}.
\par
We are going to study the system of one scalar field governed by the action
\begin{equation}
  S(\phi)={1\over 2}\int_{_{\R\times M}}d^{^{2+1}}x\left( \partial^\mu\phi^*\partial_\mu\phi+m^2|\phi|^2\right)
  +{1\over 2}\int_{_{\R\times\partial M}}d^{^{2}}x\phi^*\partial_n\phi
\end{equation}
with physical space given in figure \ref{1dwire}. From geometrical point of view the physical space where
the quantum field is confined is $M=[0,L]\times\R$, but there are two identical lattices at $x=0$ and $x=L$
respectively. This fact implies that, the boundary data of the quantum field should preserve this lattice
structure, i. e. the boundary data should have the same periodicity of the lattice. Since the two lattices
are identical, and no displacement between them appears, in this case boundary data should only have an unique
periodicity. This geometry is isomorphic to a cylindrical space-time where the longitudinal direction corresponds
to the direction orthogonal to the plates, and the circled direction is the parallel direction to the plates.
In addition, the simple periodic structure imposed implies that the field over the cylinder has winding number $1$.

In order to determine the set $\mpro_{rF}$ of regular self-adjoint extensions, we need to construct the most general
wave function for the quantum mechanical system associated to the QFT between crystal plates. Since
there is no potential between plates, and in the plates, the only requirement to be imposed, is the
periodicity of the wave function the direction parallel to the crystal (the $y$-direction):
\begin{equation}
  \psi_{n,k}(x,y)=A \e^{i (k x+q_n y)}+ B \e^{i (k x-q_n y)}+C \e^{i (-k x+q_n y)}+D \e^{-i (k x+q_n y)} \label{gwf}
\end{equation}
being $q_n =2\pi n /a$ with $n\in\mathbb{N}$. Since the momenta in the $y$-direction are discrete
(due to the requirement of periodicity imposed in order to take into account the crystal structure and
Bloch's theorem), and the x direction is compact and bounded ($x\in[0,L]$), the allowed momenta in the
$x$-direction are also a discrete set. The allowed values of momentum $k$ should arise from the boundary condition.

\section{The set of regular boundary conditions: ${\cal M}_{rF}$.}

In order to exactly determine the set $\mathcal{M}_r$ of regular boundary conditions (reduced self-adjoint
extensions), we need to explicitly compute the equations that arise from the general boundary condition
given by equation \ref{gbc}. Using same notation used in \cite{jmmcphd} the boundary data can be written as
\begin{equation}
  \varphi\pm i\dot\varphi=\Psi_\pm\equiv
  \left( \begin{array}{c}
  \psi_{kn}(0,y)\mp i\partial_x\psi_{kn}(0,y) \\
  \psi_{kn}(l,y)\pm i\partial_x\psi_{kn}(L,y) \\\end{array}\right)
\end{equation}
and the boundary condition takes the form
\begin{equation}
  \Psi_-=U\left(\Psi_+\right)
\end{equation}

From equation \ref{gwf} we can obtain an explicit expression for $\Psi_\pm$:
\begin{eqnarray}
  \psi_{k,n}(0,y)\pm i\partial_x \psi_{k,n}(0,y)&=&\nonumber{\left(A(1\pm ik)+C(1\mp ik)\right)\e^{i q_n y}}\\
  &+& \left(B(1\pm ik)+D(1\mp ik)\right)\e^{-i q_n y}\\
   \psi_{k,n}(L,y)\pm i\partial_x \psi_{k,n}(L,y)&=&\nonumber{\left(A\e^{i k L}(1\pm ik)+C\e^{-i k L}(1\mp ik)\right)\e^{i q_n y}}\\
   &+& \left(B\e^{i k L}(1\pm ik)+D\e^{-i k L}(1\mp ik)\right)\e^{-i q_n y}
\end{eqnarray}
In view of these last expressions, and taking into account that the functions $\e^{iq_n y}$ and $\e^{-iq_n y}$, are linear independent, we can rewrite all in terms of four dimensional column vectors and 4x4 matrices\footnote{$\left(\widetilde{\Psi}_{\pm}\right)_{1,3}$ are the $\e^{i q_n L}$ components, and $\left(\widetilde{\Psi}_{\pm}\right)_{2,4}$ the $\e^{-i q_n L}$ components. Also we could have imposed the convention of choosing $\left(\widetilde{\Psi}_{\pm}\right)_{1,2}$ to be the $\e^{i q_n L}$ components, and $\left(\widetilde{\Psi}_{\pm}\right)_{3,4}$ the $\e^{-i q_n L}$ components, but the spectral function does not change under this change, due to the properties of the determinant ($\det(A B)=\det(A)\det(B)$).}. The 2 dimensional column vector $\Psi_\pm$ becomes now
\begin{equation}
  \widetilde{\Psi}_{\pm}=\left(\begin{array}{c}
  A(1\pm ik)+C(1\mp ik) \\
  B(1\pm ik)+D(1\mp ik) \\
  A\e^{i k L}(1\pm ik)+C\e^{-i k L}(1\mp ik) \\
  B\e^{i k L}(1\pm ik)+D\e^{-i k L}(1\mp ik) \\
  \end{array}\right)\label{psipm}
\end{equation}
and the boundary condition is written in terms of a 4x4 unitary matrix:
\begin{equation}
  \widetilde\Psi_-(k;L)-=U\left(\widetilde\Psi_+(k;L)\right);\quad U\in {\rm U}(4)
\end{equation}
\begin{obs}
  Not all matrices $U\in{\rm U}(4)$ are allowed boundary conditions. This means, that not for all $U\in{\rm U}(4)$ the theory of quantum fields obtained is unitary, and well defined.
\end{obs}
 In order to classify which $U\in{\rm U}(4)$ give rise to unitary QFT's we must make use of the \textit{consistency lemma} (see \cite{jmmcphd}). From the \textit{consistency lemma} is rapidly deduced that the space of regular boundary conditions is given by the subset:
 \begin{equation}
   \mathcal{M}_r=\left\{U\in{\rm U}(4)|\,\,\forall\e^{i \theta}\in {\rm spec}(U), \tan \left(\frac{\theta}{2}\right)\geq 0\right\}
 \end{equation}
${\cal M}_r$ is a space with boundary, given by:
\begin{equation}
   \partial\mathcal{M}_r=\left\{U\in{\rm U}(4)|\,\,\exists\e^{i \theta}\in {\rm spec}(U), \tan \left(\frac{\theta}{2}\right)= 0\right\}
 \end{equation}
 Hence, ${\cal M}_r$ is a 15 dimensional manifold, with 14 dimensional boundary $\partial{\cal M}_r$. Since the sequence
 \begin{equation}
   1\hookrightarrow{\rm U}(1)\hookrightarrow{\rm U}(4)\rightarrow{\rm U}(1)\rightarrow 1
 \end{equation}
 is exact via the determinant projection of U(4) into U(1), we can write ${\rm U}(4)={\rm U}(1)\times {\rm SU}(4)$ to get a simplified coordinate system in ${\rm U}(4)$, and hence also for ${\cal M}_r$:
 \begin{equation}
   {\rm U}(4)=\left\{\e^{i \theta} U| U\in{\rm SU}(4),\,\e^{i\theta}\in {\rm U}(1)\right\}
 \end{equation}
 An explicit parametrization for ${\rm SU}(4)$ will be given later.

 \section{Explicit computation of the spectral function.}
 In this section we will compute explicitly the spectral function whose zeroes give the allowed momenta for the transverse direction, in terms of the boundary condition. This computation has never been done before, and is very useful from the computational point of view.
 \par
 Defining the four-vector
 \begin{equation}
   \Phi=\left(
      \begin{array}{c}
        A \\
        B \\
        C \\
        D \\
      \end{array}
    \right)
 \end{equation}
and having in mind equation (\ref{psipm}), rapidly allows us to write $\Psi_{\pm}$ as
\begin{equation}
  \Psi_\pm=M_\pm(k)\dot\Phi.
\end{equation}
The matrices $M_\pm(k)$ are defined as:
\begin{equation}
  M_\pm(k)=\left(
  \begin{array}{cccc}
  1\mp i k & 0 & 1\pm i k & 0 \\
  0 & 1\mp i k & 0 & 1\pm i k \\
  \e^{i a k} (1\pm i k) & 0 & \e^{-i a k} (1\mp i k) & 0 \\
  0 & \e^{i a k} (1\pm i k) & 0 & \e^{-i a k} (1\mp ik)
\end{array}
\right)
\end{equation}
Now the boundary condition is written in terms of $M_\pm(k)$ and $\Phi$ as
\begin{equation}
  \left[M_-(k)-U\cdot M_+(k)\right]\cdot\Phi=0\label{gbcm}
\end{equation}

\begin{obs}
  The  matrices $M_\pm(k)$ have determinant
  \begin{equation}
    D_\pm(k)=-4\left[2 k \cos(kL)\mp(k^2-1)\sin(kL)\right]^2\label{detm}
  \end{equation}
  and therefore they are singular for a discrete but infinite number of values of $k$, i. e. the equation $D_\pm(k)=0$ has an infinite number of solutions. Nevertheless, $D_\pm(k)=0$ does not determine the allowed values of $k$ because roots of $D_+(k)$ do not coincide with roots of $D_-(k)$.
\end{obs}

In order to have non-trivial solutions satisfying (\ref{gbcm}), and in view of the observation made above, we must require the condition:
\begin{equation}
   \det\left[M_-(k)-U\cdot M_+(k)\right]=0
\end{equation}
This condition, as is known from \cite{jmmcphd} gives rise to the spectral function, that is now defined as:
\begin{equation}
  h_U(k;L)\equiv\det\left[M_-(k)-U\cdot M_+(k)\right]
\end{equation}
\begin{lemma}
  The infinite discrete set of allowed values for the transverse momentum $k$ are given by the set of real positive zeros of the spectral function $h_U(k;L)$. Note that AIM theorem ensures the self-adjointness, meanwhile the consistency lemma ensures the unitarity of the QFT. This means that after imposing both conditions, the boundary condition can not give rise to bound states, i. e., these two conditions ensure that the spectral function does not have imaginary zeros.
\end{lemma}

Explicit computation of the spectral function can be very useful form a pure computational point of view. Even when the spectral function has a very large expression, is fundamentally composed by trigonometric functions with argument $k L$ and polynomials in $k$ whose coefficients depend on the boundary condition. This means that calculating its zeros numerically is not a difficult task, with most recent methods. We proceed now to give the explicit algebraic structure of the spectral function. Later on some particular cases for matriz $U$ will be studied.

For studying the algebraic structure of $h_U(k;L)$ we introduce an auxiliary real positive parameter $\epsilon$, and define
\begin{equation}
  \widetilde h_U(k;L;\epsilon)=\equiv\det\left[M_-(k)-\epsilon U\cdot M_+(k)\right].
\end{equation}
Trivially we can re-store the spectral function by making $\epsilon=1$. The key point is that $\widetilde h_U(k;L;\epsilon)$ is a four order polynomial in $\epsilon$, and the order in $\epsilon$ coincides with the number of $U$-matrix elements appearing. Hence we can rapidly identify where the algebraic invariants of matrix $U$ appear:
\begin{itemize}
  \item Terms of order 4 in $\epsilon$ are the only ones that include $\det (U)$.
  \item Terms of order 1 in $\epsilon$ are the only ones where ${\rm tr}(U)$ can appear.
  \item Terms of order 0 in $\epsilon$ do not depend on matrix $U$ and hence are independent of the boundary condition.
\end{itemize}
We introduce at this point the notation for the coefficients in $\epsilon$ of $\widetilde h_U(k;L;\epsilon)$:
\begin{equation}
  \widetilde h_U(k;L;\epsilon)=c_0(k,L)+c_1(k,L,U)\epsilon+c_2(k,L,U)\epsilon^2+c_3(k,L,U)\epsilon^3 +c_4(k,L,U)\epsilon^4
\end{equation}
Coefficients itemized before as the ones that include algebraic invariants of the $U$ matrix, will be named from now on, {\it algebraic terms of $\widetilde h_U(k;L;\epsilon)$ }.

\subsection{The algebraic terms of $\widetilde h_U(k;L;\epsilon)$.}
In this sub-section we show explicitly how the the algebraic invariants of matrix $U$ enter in the spectral function via the $\widetilde h_U$ function.

\paragraph{The coefficient $c_0(k,L)$} intuitively can only depend on the matrix $M_-$ since it is the only part of $\widetilde h_U(k;L;\epsilon)$ that does not multiplied by $U$. The explicit calculation ensures that intuition points in the right direction, and the 0 order coefficient is given by:
\begin{equation}
  c_0(k,L)=\det \left(M_-(k,L)\right)=-4\left[2 k\cos(kL)+(k^2-1)\sin(kL)\right]^2
\end{equation}

\paragraph{The coefficient $c_1(k,L,U)$} is the one where ${\tr }(U)$ can appear. In general also non-diagonal terms can appear. After calculation, the final expression for the coefficient $c_1(k,L,U)$ is:
\begin{eqnarray}
  c_1(k,L,U)&=&\nonumber{\left(2 k \cos (kL)+(k^2-1)\sin (kL)\right)\times}\\
  &&\left(-4(k^2+1){\rm tr}(U)\sin (kL)+8k \left(U_{13}+U_{31}+U_{24}+U_{42}\right)\right) \label{coef1}
\end{eqnarray}
As can be rapidly seen, $c_1(k,L,U)$ depends on ${\rm tr}(U)$ but also has a non-algebraic part depending on non-diagonal matrix elements.
\paragraph{The coefficient $c_4(k,L,U)$} is the one where the determinant appears. Since it comes from the fourth order in the $\epsilon$ expansión, in this case no non algebraic terms can appear, since all combinations of four matrix elements of $U$ appear in $\det(U)$. After calculation and simplification, the corresponding coefficient reads:
\begin{equation}
  c_4(k,L,U)=\det(U)det(M_+)=-4\det(U)\left[2k\cos (kL)-(k^2-1)\sin(kL)\right]^2
\end{equation}

\subsection{The non-algebraic terms of $\widetilde h_U(k;L;\epsilon)$.}
Calculation of the non-algebraic terms is a more tedious than for the algebraic ones. Nevertheless they can also be computed using standard simplification techniques.
\paragraph{The second order coefficient $c_2(k,L,U)$} contains only combinations of two matrix elements of matrix $U$. We could in general expect all possible combinations to appear. After calculation and simplification the coefficient is given by:

\begin{eqnarray}
 c_2(k,L,U)&=&\nonumber{16 k^2 \left(U_{14} U_{23}-U_{13} U_{24}-U_{24}
   U_{31}+U_{21} U_{34}+U_{32} U_{41}+U_{13}
   U_{42}-U_{31} U_{42}+U_{12} U_{43}\right)}\\
   &+&\nonumber{4 \left(k^2+1\right)^2 \sin ^2(k L) \left(U_{12} U_{21}-U_{11}
   U_{22}+U_{23} U_{32}-U_{22} U_{33}\right.}\\
   &+&\nonumber{\left.U_{14}
   U_{41}+U_{34} U_{43}+U_{11} U_{44}-U_{33}
   U_{44}\right) }\\
   &+&\nonumber{\left(16 k^2-4 \left(k^2+1\right)^2 \sin ^2(k L)\right)\left(U_{11} U_{33}-U_{13} U_{31}
   +U_{22} U_{44}-U_{24}U_{42}\right) }\\
   &-&\nonumber{8 k \left(k^2+1\right) \sin (k L) \left(U_{12} U_{23}+U_{34}
   U_{23}-U_{11} U_{24}-U_{22} U_{31}\right.}\\
   &+&\nonumber{U_{21}U_{32}-U_{24} U_{33}+U_{12} U_{41}+U_{34}U_{41}-U_{11} U_{42}-U_{33} U_{42}
   +U_{32}U_{43}+U_{14}U_{21}}\\
   &+&\left.U_{14}U_{43}-U_{31}U_{44}-U_{13}U_{22}+U_{13}U_{44}\right)
\end{eqnarray}

\paragraph{The coefficient $c_3(k,L,U)$} is obtained after a similar calculation of the one made for $c_2$:
\begin{eqnarray}
  c_3(k,L,U)&=&\nonumber{-4(k^2+1)\sin(kL)\left((k^2-1)\sin(kL)+2k\cos(kL)\right)\left[U_{11}U_{22}U_{33}\right.}\\
  &+&\nonumber{U_{12}U_{23}U_{31}+U_{13}U_{32}U_{21}-U_{11}U_{32}U_{23}- U_{13}U_{22}U_{31}-U_{12}U_{21}U_{33}}\\
  &+&\nonumber{U_{11}U_{22}U_{44}+U_{12}U_{24}U_{41}+U_{21}U_{42}U_{14}-U_{11}U_{24}U_{42} -U_{14}U_{22}U_{41}}\\
  &-&\nonumber{U_{12}U_{21}U_{44}+U_{11}U_{33}U_{44}+U_{13}U_{34}U_{41}+U_{31}U_{43}U_{14}-U_{11}U_{34}U_{43}}\\
  &-&\nonumber{U_{14}U_{33}U_{41}-U_{13}U_{31}U_{44}+U_{22}U_{33}U_{44}+U_{23}U_{34}U_{42}+U_{32}U_{43}U_{24}}\\
  &-&\nonumber{\left.U_{22}U_{34}U_{43} -U_{24}U_{33}U_{42}-U_{23}U_{32}U_{44}\right]}\\
  &-&\nonumber{8k\left(2k\cos(kL)-(k^2-1)\sin(kL)\right)\left[U_{11}U_{33}U_{24}+U_{11}U_{33}U_{42}\right.}\\
  &+&\nonumber{U_{13}U_{32}U_{41}+U_{13}U_{34}U_{21}+ U_{31}U_{12}U_{43}+U_{31}U_{14}U_{23}-U_{11}U_{23}U_{34}}\\
  &-&\nonumber{U_{12}U_{33}U_{41}-U_{11}U_{32}U_{43}-U_{21}U_{33}U_{14}- U_{13}U_{31}U_{42}-U_{13}U_{31}U_{24}}\\
  &+&\nonumber{U_{22}U_{44}U_{13}+U_{22}U_{44}U_{31}+U_{24}U_{41}U_{32}+U_{42}U_{21}U_{34} +U_{24}U_{43}U_{12}} \\
  &+&\nonumber{U_{42}U_{23}U_{14}-U_{22}U_{14}U_{43}-U_{22}U_{34}U_{41}-U_{12}U_{44}U_{23}-U_{32}U_{44}U_{21}}\\ &-&\left.U_{24}U_{42}U_{13}-U_{24}U_{42}U_{31}\right]
\end{eqnarray}
Although $c_3(k,L,U)$ can be though to be constructed in terms of non algebraic invariants of matrix $U$, it can be rewritten in terms of algebraic invariants of a matrix associated to the matrix $U$: the matrix of minors. Also this way of writing $c_3$ provides a way to simplify a lot the expression for $c_3$. First of all recall the definition of the matrix of minors.In the most general case, if $U$ is a square $n\times n$ matrix the $n^2$ $(n-1)\times (n-1)$ submatrices are constructed deleting one row and one column, and we use the notation:
\begin{equation}
  U^{(i,j)}\equiv\left\{{\rm delete\,\,row}\,\, n+1-i,\,\,{\rm and\,\,column}\,\, n+1-j\right\}
\end{equation}
With this notation, we define the $n\times n$ matrix of minors $A^{(U)}$ associated to $U$, as the matrix whose elements $a^{(U)}_{ij}$ are given by:
\begin{equation}
  a^{(U)}_{ij}\equiv\det\left(U^{(i,j)}\right)
\end{equation}
Using the matrix of minors associated to $U$, the coefficient $c_3$ can be written in a very compact and symmetric way:
\begin{eqnarray}
  c_3(k,L,U)&=&\nonumber{\left(2k\cos(kL)-(k^2-1)\sin(kL)\right)\left[4(k^2+1)\sin(kL){\rm tr}\left(A^{(U)}\right)\right.}\\
  &+&\left.8k\left(a^{(U)}_{13}+a^{(U)}_{31}+a^{(U)}_{24}+a^{(U)}_{42}\right)\right]
\end{eqnarray}

After this simplification, we have explicitly computed the expression for the spectral function whose zeros are the allowed values for the transverse momentum:
\begin{equation}
  h_U(k;L)=c_0(k,L)+c_4\left(k,L,\det(U)\right)+c_3\left(k,L,A^{(U)}\right)+c_2(k,L,U)
\end{equation}

\section{Particular cases.}
\subsection{Diagonal, and anti-diagonal elements of ${\cal M}_r$.}
\paragraph{Diagonal elements of ${\cal M}_r$} are given by four parameters, that are their eigenvalues:
\begin{equation}
  U_d={\rm diag}\left(\lambda_1,...,\lambda_4\right)
\end{equation}
Each eigenvalue is a modulus 1 complex number $\lambda_j=e^{i\theta_j}$, satisfying the consistency condition $\tan\left(\theta_j/2\right)\geq0$. In this case the corresponding minors matrix $A^{(U_d)}$ is also diagonal, and given by:
\begin{equation}
  A^{(U_d)}={\rm diag}\left(\lambda_1\lambda_2\lambda_3,\, \lambda_1\lambda_2\lambda_4,\, \lambda_1\lambda_3\lambda_4,\, \lambda_2\lambda_3\lambda_4\right)
\end{equation}
If we define the two variable polynomial
\begin{equation}
p(x,y;k,L)=2 i\left(k^2+1\right) \sin (k L) (x+y) -(D_+)^{1/2}x y+(D_-)^{1/2}
\end{equation}
the corresponding spectral function can be written in a compact form as:
\begin{equation}
  h_{U_d}(k,L)=p(\lambda_1,\lambda_3; k,L)p(\lambda_2,\lambda_4; k,L)
\end{equation}
There are some particular cases of diagonal boundary conditions that should be explicitly computed. The two most important cases are Dirichlet and Neumann boundary conditions, given by:
\begin{equation}
  U_D=-\I;\quad U_N=\I
\end{equation}
The spectral function for both cases looks very simple, and its roots can be calculated analytically:
\begin{equation}
  h_D(k,L)=-64 \sin^2(kL),\quad h_D(k,L)=-64 k^4 \sin^2(kL)
\end{equation}
Also we can mix Neumann and Dirichlet boundary conditions, by imposing one boundary condition for the upwards modes ($\e^{iq y}$ components), and the other one for the downards modes ($\e^{-iq y}$ components), which is given by a diagonal matrix
\begin{equation}
  U_{ND}\left(
     \begin{array}{cc}
       -\I & 0 \\
       0 & \I \\
     \end{array}
   \right)
\end{equation}
In this case the spectral function resulting is also very simple, and given by
\begin{equation}
  h_{ND}(k,L)=-64 k^2 \cos^2(a k)
\end{equation}

\paragraph{Anti-diagonal elements of ${\cal M}_r$} can be handled with analytical techniques. Let us write an arbitrary anti-diagonal element of ${\cal M}_r$ as
\begin{equation}
  U_{ad}=\left(
          \begin{array}{cccc}
            0 & 0 & 0 & \lambda_1 \\
            0 & 0 & \lambda_2 & 0 \\
            0 & \lambda_3 & 0 & 0 \\
            \lambda_4 & 0 & 0 & 0 \\
          \end{array}
        \right).
\end{equation}
The corresponding matrix of minors is easily computed, and has the form:
\begin{equation}
  A^{(ad)}=\left(
          \begin{array}{cccc}
            0 & 0 & 0 & -\lambda_2\lambda_3\lambda_4 \\
            0 & 0 & -\lambda_1\lambda_3\lambda_4 & 0 \\
            0 & -\lambda_1\lambda_2\lambda_4 & 0 & 0 \\
            -\lambda_1\lambda_2\lambda_3 & 0 & 0 & 0 \\
          \end{array}
        \right).
\end{equation}
Simplifying the general expression for the spectral function in the case of anti-diagonal boundary conditions, we obtain the spectral function
\begin{equation}
  h_{ad}(k,L)=D_-+D_+\lambda_1\lambda_2\lambda_3\lambda_4+4(k^2+1)^2\sin^2(kL)(\lambda_2\lambda_3+\lambda_1\lambda_4) +16k^2(\lambda_1\lambda_2+\lambda_3\lambda_4)
\end{equation}
Well known boundary conditions belong to this group. The most important ones are periodic, and anti-periodic boundary conditions. For the periodic boundary conditions the corresponding matrix is
\begin{equation}
  U_p=\left(
          \begin{array}{cccc}
            0 & 0 & 0 & 1 \\
            0 & 0 & 1 & 0 \\
            0 & 1 & 0 & 0 \\
            1 & 0 & 0 & 0 \\
          \end{array}
        \right)
\end{equation}
and the spectral function obtained has the form:
\begin{equation}
  h_p(k,L)=64 k^2 \sin^2(k L)
\end{equation}
For the case of anti-periodic boundary conditions, the corresponding matrix is $U_{ap}=-U_p$, which leads to the same spectral function as for periodic boundary conditions: $h_{ap}(k,L)=h_p(k,L)$. We can also mix periodic and anti-periodic boundary conditions making the system to distinguish between upwards propagating modes, and downwards propagating modes:
\begin{equation}
  U_{pap}=\left(
          \begin{array}{cccc}
            0 & 0 & 0 & 1 \\
            0 & 0 & 1 & 0 \\
            0 & -1 & 0 & 0 \\
            -1 & 0 & 0 & 0 \\
          \end{array}
        \right).
\end{equation}
The spectral function generated in this case is given by
\begin{equation}
  h_{pap}(k,L)=-16 (-1 + k^2)^2 \sin^2(k L).
\end{equation}
For this case is immediately observed from the spectral function that there appears a doubly degenerated state for $k=1$ which does not depend on the distance $L$ between crystal wires, and hence does not contribute to the vacuum force between plates.

\subsection{Box-anti-diagonal elements of ${\cal M}_r$.}
Other well known important boundary conditions, studied for the case of homogeneous isotropic plates, arise in the case of one dimensional perfect lattices, as box-anti-diagonal boundary conditions, i.e. boundary conditions given by matrices of the form:
\begin{equation}
  U=\left(
      \begin{array}{cc}
        0 & A \\
        B & 0 \\
      \end{array}
    \right)
\end{equation}
where $A, B\in{\rm U} (2)$. This subset of boundary conditions is contained in $\left({\rm U} (2)\times{\rm U} (2)\right)\bigcap{\cal M}_{rF}$. The two most well-known boundary conditions belonging to this subset are the pseudo-periodic and quasi-periodic boundary conditions. Both examples correspond to one-parameter set of boundary conditions that interpolate between periodic and anti-periodic boundary conditions.

\paragraph{The pseudo-periodic boundary conditions} physically correspond to a situation in which both plates are identified (cylinder geometry), with a magnetic flux passing along the cylinder. This situation can be developed with a boundary condition given by the box diagonal matrix
\begin{equation}
  U_{pp}=\left(
\begin{array}{cccc}
 0 & 0 & 0 & e^{-i \alpha } \\
 0 & 0 & e^{i \alpha } & 0 \\
 0 & e^{i \alpha } & 0 & 0 \\
 e^{-i \alpha } & 0 & 0 & 0
\end{array}
\right)
\end{equation}
It must be pointed, that if the upwards modes are suffering a flux $\alpha$ then, the downwards modes coupled to the same flux are suffering a flux $-\alpha$, as is perfectly shown in the corresponding unitary matrix. The resulting spectral function is given by
\begin{equation}
  h_{pp}(k,L)=8 (-1 + 6 k^2 - k^4 + (1 + k^2)^2 \cos(2 \alpha)) \sin^2(k L)
\end{equation}
If now we couple downwards and upwards modes to different fluxes, the corresponding boundary condition is given by
\begin{equation}
  U_{mpp}=\left(
\begin{array}{cccc}
 0 & 0 & 0 & e^{-i \alpha } \\
 0 & 0 & e^{i \alpha } & 0 \\
 0 & e^{i \beta } & 0 & 0 \\
 e^{-i \beta } & 0 & 0 & 0
\end{array}
\right),
\end{equation}
which leads to the spectral function
\begin{equation}
  h_{mpp}(k,L)=8 (-1 + 6 k^2 - k^4 + (1 + k^2)^2 \cos(\alpha + \beta)) \sin^2(k L)
\end{equation}
As expected, when $\beta=\alpha$ the periodic boundary conditions are re-stored, because the flux suffered by upwards modes, is compensated by the effect of the flux suffered by the downwards modes.

\paragraph{Quasi-periodic boundary conditions} are mathematically very similar to pseudo-periodic boundary conditions, but the corresponding physical situation is very different. Quasi-periodic boundary conditions correspond to an identification of both plates via a delta function, i.e an step discontinuity in the first derivatives at the junction point. These boundary conditions are given by the matrix

\begin{equation}
  U_{qp}=\left(
\begin{array}{cccc}
 0 & 0 & \cos (\alpha ) & \sin (\alpha ) \\
 0 & 0 & \sin (\alpha ) & -\cos (\alpha ) \\
 \cos (\alpha ) & -\sin (\alpha ) & 0 & 0 \\
 -\sin (\alpha ) & -\cos (\alpha ) & 0 & 0
\end{array}
\right)
\end{equation}
and the corresponding spectral function reads:
\begin{equation}
  h_{qp}(k,L)=8 (-1 + 6 k^2 - k^4 + (1 + k^2)^2 \cos(2 \alpha) \sin^2(k L).
\end{equation}

\section{Conclusions.}
In the first part of the paper we classified the set boundary conditions that give rise to a unitary massless scalar quantum field theory with regular behavior of the field in the boundary (given by the two one dimensional crystal plates, or equivalently by two $S^1$ circles).In the second part we determined uniquely the spectrum of $\ol_u$ by calculating the spectral function for each self adjoint extensión $\ol_U\in \mpro_{rF}$. The importance of the spectral function in relation to Casimir effect is that having the spectral function explicitly computed it can be computed the Casimir energy as a global function over the 16 dimensional space $\mpro_{rF}$. Taking into account that the vacuum energy is given basically by
\begin{equation}
  E_c(U)=\sum_{n\in\N}\sum_{k\in Z\left(h_u\right)}\sqrt{k^2+\frac{4\pi^2n^2}{a^2}}
\end{equation}
and the fact that the spectral function $h_U(k)$ is holomorphic in $k$, we can replace the summation in $k$ by an integration in the complex $k$-plane over a contour $C$ that encloses all the zeroes of $h_u(k)$:
\begin{equation}
  \sum_{k\in Z\left(h_u\right)}f(k)\longrightarrow\oint_C\frac{dk}{2\pi i}f(k)\frac{d }{dk}\ln\left(h_U(k)\right)
\end{equation}
Hence using the heat kernel regularization
\begin{equation}
  E_c(U;\epsilon)=\sum_{n\in\N}\sum_{k\in Z\left(h_u\right)}\sqrt{k^2+\frac{4\pi^2n^2}{a^2}} \e^{-\epsilon\sqrt{k^2+\frac{4\pi^2n^2}{a^2}}}
\end{equation}
in the same way as done in \cite{jmmcphd} the spectral function allows to obtain a computable (numerically in general and analytically in particular simple cases) expression for the vacuum energy and the corresponding force.
\par
Finally in the third part of the paper, we have computed explicitly the spectral function for the most well-know types of boundary conditions, as well as some generalizations to one-parameter and two-parameter families of boundary conditions.

\section*{Acknowledgement}
The authors benefited from exchange of ideas by the ESF Research Network
CASIMIR. JMMC acknowledges support from DFG, grant number BO 1112/19-1.

\end{document}